\documentclass[12pt]{article}
\usepackage{amsfonts,amsmath,amssymb}
\usepackage{graphicx,color}
\newcommand{\be}{\begin{equation}}
\newcommand{\ba}{\begin{eqnarray}}
\newcommand{\ea}{\end{eqnarray}}
\newcommand{\ee}{\end{equation}}

\begin{document}

\begin{flushright} 
December 2008  \\
KUNS-2172\\
WIS/22/08-DEC-DPP\\
RIKEN-TH-145\\
APCTP Pre2008-010
\end{flushright} 

\vspace{0.1cm}

\begin{center}
  {\LARGE 
  Worldsheet Analysis of Gauge/Gravity Dualities
  }
\end{center}
\vspace{0.1cm}
\begin{center}

         Tatsuo A{\sc zeyanagi}$^{a}$\footnote
           {
E-mail address : aze@gauge.scphys.kyoto-u.ac.jp}, 
         Masanori H{\sc anada}$^{b}$\footnote
         {
E-mail address : masanori.hanada@weizmann.ac.il},    
         Hikaru K{\sc awai}$^{a,c}$\footnote
           {
E-mail address : hkawai@gauge.scphys.kyoto-u.ac.jp} and 
         Yoshinori M{\sc atsuo}$^{d}$\footnote
         {
E-mail address : ymatsuo@apctp.org}         

\vspace{0.3cm}

${}^{a}$
{\it Department of Physics, Kyoto University,\\
Kyoto 606-8502, Japan}\\

${}^{b}$
{\it Department of Particle Physics, Weizmann Institute of Science,\\
     Rehovot 76100, Israel }\\

${}^{c}$
{\it Theoretical Physics Laboratory, Nishina Center, RIKEN,\\
 Wako, Saitama 351-0198, Japan}\\

${}^{d}$
{\it Asia Pacific Center for Theoretical Physics,\\
Pohang, Gyeongbuk 790-784, Korea}\\
\end{center}

\vspace{1.5cm}

\begin{center}
  {\bf abstract}
\end{center}

Gauge/gravity dualities are investigated 
from the worldsheet point of view. 
In \cite{KS07}, a 
duality between 4d SYM   
and  
supergravity on $AdS_5\times S^5$ 
has been partly explained 
by using an anisotropic scale invariance of worldsheet theory. 
In this paper, we refine the argument and 
generalize it to lower dimensional cases. 
We show the correspondence 
between the Wilson loops in $(p+1)$-d SYM and 
the minimal surface in the black $p$-brane background.  
Although the scale invariance does not exist in these cases, 
the generalized scale transformation can be utilized. 
We also find that the energy density of open strings 
can be related to the ADM mass of the $p$-brane 
without relying on this symmetry.

\newpage
\newpage
\section{Introduction}
Since the AdS/CFT correspondence has been conjectured \cite{Maldacena98},  
many evidences have been found 
and various applications have been proposed by assuming 
its correctness.   
However, it is still not clear to what extent this correspondence holds. 

As a concrete example, let us consider the original $AdS_5\times S^5$ setup. 
In the narrowest sense, it is conjectured that 4d ${\cal N}=4$ super Yang-Mills 
theory (SYM) is equivalent to type IIB supergravity 
on the $AdS_5\times S^5$ background, 
in the large-$N$ and strong 't Hooft coupling limit.  
From various nontrivial checks, 
the conjecture seems to be correct at least in this limit. 
This correspondence is often extended to 
finite 't Hooft coupling, for which 
the SYM corresponds to type IIB string theory 
in the classical limit. 
In some cases, correspondence 
between finite-$N$ SYM and fully quantized type IIB string is considered. 
However, it is not apparent 
whether it really holds in such a wider sense. 
It is also not clear whether 
the duality between theories with lower symmetries can hold. 
Clarification of such issues is crucial for correct application of the duality. 
String worldsheet viewpoint is expected to shed light on these questions  
and may provide a {\it proof} of the duality. 

In \cite{KS07}, a part of 
the narrowest correspondence is explained 
by considering worldsheets of strings 
propagating in the background of $N$ coincident D3-branes
\footnote{For another interesting approach from worldsheet viewpoint, 
see \cite{BV07}.}.
The essential point is that an anisotropic 
scale invariance holds as long as worldsheets are located in the 
near horizon region. This symmetry can be used 
to directly relate SYM to supergravity.  
If we introduce a source D-brane close to the D3-branes, 
the string worldsheets stretched between them can be 
regarded as a Wilson loop in $\mathcal{N}=4$ $SU(N)$ SYM. 
By applying the scale transformation so that
the source D-brane is distant from the D3-branes
but still in the near horizon region,
this system is described by 
supergravity on the $AdS_5\times S^5$ background. 
In this way, SYM and supergravity are related to each other by the scale transformation. 

In \cite{IMSY98}, it is proposed that 
there are dualities between $(p+1)$-d SYM and 
type II string theory on black $p$-brane background for $p\neq 3$. 
At large-$N$ and strong 't Hooft coupling, the energy density and  
Wilson loop in SYM are conjectured to correspond to the ADM mass 
and a minimal surface of string worldsheet in the gravity side \cite{WilsonLoop}. 
In this case, SYM is not conformal and 
the background in the gravity side is not AdS. 
However, recent numerical simulations for $p=0$ SYM  
\cite{AHNT07,CW08,HMNT08,HHNT08} 
support the validity of the correspondence.

In this paper, we refine the proposal of \cite{KS07} and  
generalize it to lower dimensional D$p$-branes 
and of finite temperature. 
We notice that the near horizon geometries of these  
branes do not have the scale transformation as an isometry. 
To see the correspondence of the energy density, 
it turns out that no symmetries are needed. 
By examining the origin of the ADM mass, 
we can show that it is exactly equal to the energy 
of the D-brane system. In the near extremal limit, 
the system is described by SYM.
Then, by subtracting the D-brane tension from the ADM mass, 
we obtain the energy density of SYM. 

\begin{figure}[tb]
\begin{center}
\includegraphics[scale=3.5]
                {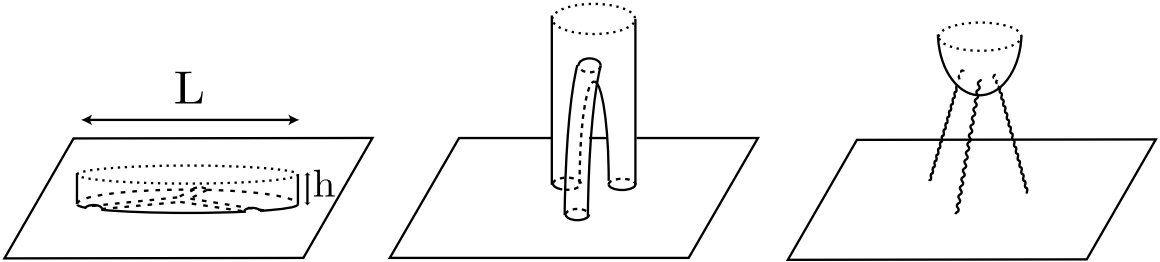}
\caption{The Wilson loop in SYM is represented by the string 
worldsheets stretching between a source D-brane and the $N$ coincident D-branes (left). 
By the generalized scale transformation, the source D-brane is lifted away from the
$N$ D-branes but still in the near horizon region (middle). 
This diagram can be evaluated as a minimal surface in a curved 
background (right).  
}
\label{wilsontrans}
\end{center}
\end{figure}

For the Wilson loops, a certain symmetry is needed to 
relate the gravity side to SYM side. 
For this purpose we use the generalized scale transformation 
proposed in \cite{JKY98}.
The near horizon geometry of the  
black $p$-brane \cite{HS91} in the unit $\alpha'=1$ is given by 
\footnote{Here we assume $\lambda\gg1$, and consider the region 
$1\ll U\ll (d_p\lambda)^{\frac{1}{7-p}}$. } 
\begin{equation}
ds^{2} = \Bigl(\frac{U^{7-p}}{d_p\lambda}\Bigr)^{\frac{1}{2}}
\Bigl[\Bigl(1-\Bigl(\frac{U_0}{U}\Bigr)^{7-p}\Bigr)
dx_{p+1}^2+\displaystyle{\sum_{a=1}^{p}}dx_{a}^2\Bigr] 
+\Bigl(\frac{d_p\lambda}{U^{7-p}}\Bigr)^{\frac{1}{2}}\Bigl[ 
\frac{dU^2}{1-\Bigl(\frac{U_0}{U}\Bigr)^{7-p}}
+U^2d\Omega_{8-p}^2\Bigr]. 
\end{equation}
This geometry is invariant under the following transformation; 
\begin{eqnarray}
x_A\ &\to& c^{-1}x_A \quad(A=1,2,\cdots,p,p+1), \\
U \ &\to& cU, \\
\lambda\ &\to& c^{3-p} \lambda,     
\end{eqnarray}
where $c$ is a real and positive parameter.  
As we will see, in the near horizon region, the generalized 
scale transformation turns out to be a stringy symmetry. 
We then consider a configuration in which the source
D-brane is widely extended and located close to the D$p$-branes,
as shown in the left picture in figure~\ref{wilsontrans}.
In this case the Wilson loop is described by the low
energy effective theory of the open strings, that is, SYM.
Then we apply the generalized scale transformation 
so that the worldsheet is stretched in the vertical direction
and shrunk in the horizontal direction,
as in the middle picture in figure~\ref{wilsontrans}. 
Then, as in the right picture, 
the amplitude can be evaluated as the area of the minimal surface 
in the black $p$-brane background. 
If we locate the source D-brane sufficiently close to the D$p$-branes
in the original configuration, 
we can keep it in the near horizon region
during the deformation.
In this way, the gauge/gravity correspondence follows simply,
once we can show the invariance.

This paper is organized as follows. 
In section \ref{free_energy}, we examine 
the correspondence between the energy density of SYM and 
the ADM mass of the black $p$-brane without using any symmetries. 
In section \ref{wilson_loop}, 
we discuss the generalized scale transformation and show 
the correspondence between the Wilson loop in SYM and the minimal surface 
in supergravity by applying the transformation. 
Section \ref{conclusions_and_discussions} is devoted to 
conclusions and discussions.
\section{Energy Density and ADM Mass}
\label{free_energy}
In this section we discuss the correspondence of the 
entropy. We show it by directly comparing the energy density
of the D$p$-branes and the ADM mass in the asymptotically
flat region. We do not rely on any symmetry here. It is 
in contrast to the case of 
the Wilson loops which we will discuss in the next section.

We start with considering the system of open strings 
on the D$p$-branes at finite temperature. We do not 
impose a constraint that it is reduced to SYM. 
If we observe the system at a point distant from the 
D$p$-branes, the space-time is almost flat and
described by supergravity.
However, if we come closer to the D$p$-branes, 
in principle two types of corrections will come in.
One is the quantum gravity effect and the other is 
the $\alpha'$ correction or stringy effect. 
We suppress them by imposing the condition
\begin{eqnarray}
\beta \gg 1, \quad \lambda\gg 1,
\quad \text{and} \quad N\to\infty, 
\label{limit_gravity}
\end{eqnarray} 
where $\beta$ is the inverse temperature.
The large-$N$ limit, $N\to\infty$ with $\lambda=(2\pi)^{p-2}g_sN$ 
\footnote{In terms of SYM coupling constant $g_{YM}$, $\lambda=g_{YM}^2N$.} 
kept fixed, 
suppresses closed string loops, and the conditions 
$\lambda\gg 1$ and $\beta\gg 1$ guarantee that the 
curvature is small everywhere outside the event horizon.

\begin{figure}[t]
\begin{center}
\includegraphics[scale=3.5]
                {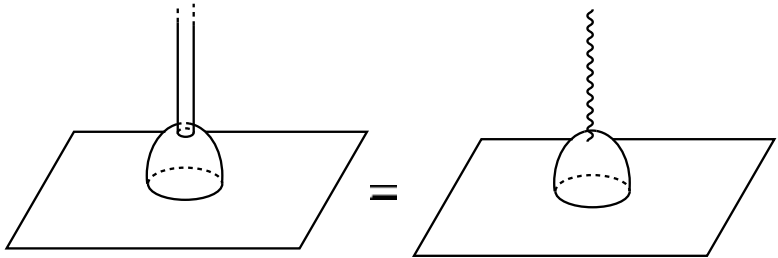}
\caption{A disk amplitude (left).
We attached a long thin tube to the worldsheet.}
\label{tube1}
\end{center}
\end{figure}

\begin{figure}[t]
\begin{center}
\includegraphics[scale=3.5]
                {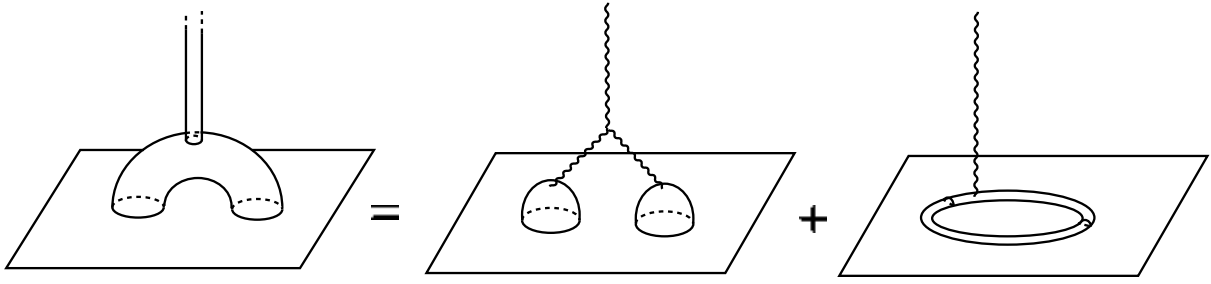}
\caption{A cylinder amplitude 
(left) are regarded as two diagrams: a diagram 
corresponding to the mass of the D$p$-branes (middle);
a one-loop diagram corresponding to 
the energy density of the open strings (right). }
\label{tube2}
\end{center}
\end{figure}

In general, the ADM mass is obtained by observing 
the gravitational field in the asymptotically flat region.
We regard the deviation of the metric from the 
flat one as a graviton in the weak field expansion, and 
the ADM mass is calculated from the linearized
Einstein equation assuming that the source of the graviton 
is the energy-momentum tensor. 
In order to apply this procedure, we first attach a tube of 
a closed string to the open string worldsheet
so that we can observe the system 
from the asymptotically flat region  
(see, for example, the left diagram of figures \ref{tube1} and \ref{tube2}). 
When the observer is distant from the D$p$-branes,
this tube becomes propagation of massless bosons,
and represents the deviation of the metric from the flat space.

Let us consider a few examples.  
First we consider the contribution from the disk diagram
(left diagram of Figure \ref{tube1}).
If we observe it in the asymptotically flat region, 
the amplitude is replaced by the right diagram 
in figure~\ref{tube1}, that is, a disk amplitude 
with a vertex insertion.  
This amplitude is independent of the temperature 
because it does not wrap on the temporal circle, 
and is proportional to $\frac{1}{U^{7-p}}$, 
where $U$ stands for the location of the observer.  
($U^{-8}$ from a propagator and $U^{p+1}$ from an integral 
over the location of the disk.) 

The second simplest diagram is the cylinder, 
which is regarded as a sum 
of two amplitudes as shown in figure~\ref{tube2}.
The middle diagram is similar to that appeared in 
the previous example, while the right is 
the one-loop vacuum diagram of open strings with 
an emission vertex attached. 
The middle diagram is independent of the
temperature and behaves as $\frac{1}{U^{2(7-p)}}$
($U^{-8\times3}$ from three propagators, $U^{-2}$ from 
a three point vertex in the bulk, $U^{10}$ from 
the location of the three point vertex and $U^{(p+1)\times 2}$ 
from two integrals over the location of the disks). 
The right one behaves as $\frac{1}{U^{7-p}}$ and 
depends on temperature. Especially it vanishes at 
zero temperature as a result of supersymmetry.
We can discuss worldsheets with more boundaries 
in a similar manner.

Because the solution of the supergravity
is uniquely determined by the mass, RR charge and isometry,
the sum of these contributions should reproduce 
the black $p$-brane solution
\begin{eqnarray}
ds^2&=&f_p^{-\frac{1}{2}}
\Bigl[\Bigl(1-\Bigl(\frac{U_0}{U}\Bigr)^{7-p}\Bigr)dx_{p+1}^2+
\displaystyle{\sum_{a=1}^{p}}dx_a^2\Bigr]+
f_p^{\frac{1}{2}}
\Bigl[\frac{dU^2}{1-(\frac{U_0}{U})^{7-p}}+U^2d\Omega_{8-p}^2\Bigr],
\nonumber\\
\label{metric_finite} \\
e^{\phi}&=&g_{s}f_p^{\frac{3-p}{4}}, 
\label{dilaton_zero}
\\
A_{1,\cdots,p,p+1}&=&\frac{1}{2g_s}(f_p^{-1}-1), 
\end{eqnarray}
where
\begin{eqnarray}
f_p=1+\frac{\alpha_p d_p\lambda}{U^{7-p}},&& \label{warpfactor_finite}\\
\alpha_p = \sqrt{1+\Bigl(\frac{U_0^{7-p}}{2d_p\lambda}\Bigr)^2}
-\Bigl(\frac{U_0^{7-p}}{2d_p\lambda}\Bigr), &&
d_p = 2^{7-p}\pi^{\frac{9-3p}{2}}\Gamma\Bigl(\frac{7-p}{2}\Bigr).
\end{eqnarray}
Here $x_{p+1}$ is the Euclidean time,
and compactified as $x_{p+1}\sim x_{p+1}+\beta$.

For example, the $(p+1,p+1)$-component of (\ref{metric_finite}) 
is expanded as 
\begin{eqnarray}
g_{p+1{\ } p+1}= 1-\left(\frac{1}{2}\alpha_p d_p\lambda
-U_0^{7-p}\right)\frac{1}{U^{7-p}}+\mathcal{O}
\Bigl(\frac{1}{U^{2(7-p)}}\Bigr).
\label{g00}
\end{eqnarray}
The right diagrams in figure \ref{tube1} and
in figure \ref{tube2} represent
parts of the second term, while
the middle diagram in figure \ref{tube2} corresponds to 
the higher order terms.

In other words, 
the sum of the worldsheet amplitudes converges for 
large enough $U$, and can be analytically continued to smaller
$U$ as (\ref{metric_finite}).
However, there are two possibilities about
the event horizon.
\begin{enumerate}
\item The D$p$-branes form a black $p$-brane.
In this case, the horizon radius $U_0$ is determined by $\beta$
from the absence of conical singularity as usual:
\begin{eqnarray} 
\beta=\frac{4\pi U_0}{(7-p)}
\Bigl(
1+\frac{\alpha_pd_p\lambda}{U_0^{7-p}}
\Bigr)^{\frac{1}{2}}.
\label{inverse_hawking_temp}
\end{eqnarray}
\item The D$p$-branes form an extended object, and
we have no event horizon. In this case we can trust the 
solution (\ref{metric_finite}) only outside the object,
and a dynamical analysis of the D$p$-branes is needed
to determine  $U_0$.
\end{enumerate}

In any case, the ADM mass is determined by the asymptotic 
$1/U^{7-p}$ behavior of the fields.
We formally integrate the $(p+1,p+1)$-component
of the energy momentum tensor $T_{\mu\nu}$ and extract the
surface term at infinity. Here  $T_{\mu\nu}$  is defined by
the weak field expansion of the Einstein equation:
\begin{eqnarray}
T_{\mu\nu} = \frac{1}{2\kappa^2}
\Bigl(\partial_{\mu}\partial_{\alpha}h^{\alpha}_{{\ }\nu}&+&
\partial_{\nu}\partial_{\alpha}h^{\alpha}_{{\ }\mu}
-\partial_{\alpha}\partial^{\alpha}h_{\mu\nu}
-\partial_{\mu}\partial_{\nu}h^{\alpha}_{{\ }\beta} \nonumber\\
&-&\eta_{\mu\nu}(\partial_{\alpha}\partial_{\beta}h^{\alpha\beta}-
\partial_{\alpha}\partial^{\alpha}h^{\beta}_{{\ }\beta})
+4(\partial_{\mu}\partial_{\nu}\phi-\eta_{\mu\nu}
\partial_{\alpha}\partial^{\alpha}\phi)
  \Bigr).\label{TmnStringFrame}
\end{eqnarray}  
In the Einstein frame $h_{\mu\nu}^{E}=h_{\mu\nu}-\frac{1}{2}\eta_{\mu\nu}\phi$,
this becomes
\begin{equation}
T_{\mu\nu} = \frac{1}{2\kappa^2}
\Bigl(\partial_{\mu}\partial_{\alpha}{h^E}^{\alpha}_{{\ }\nu} + 
\partial_{\nu}\partial_{\alpha}{h^E}^{\alpha}_{{\ }\mu}
-\partial_{\alpha}\partial^{\alpha}{h^E}_{\mu\nu}
-\partial_{\mu}\partial_{\nu}{h^E}^{\alpha}_{{\ }\beta}
- \eta_{\mu\nu}(\partial_{\alpha}\partial_{\beta}{h^E}^{\alpha\beta}-
\partial_{\alpha}\partial^{\alpha}{h^E}^{\beta}_{{\ }\beta}) \Bigr).
\end{equation} 
The ADM mass (per unit volume of the D$p$-branes) thus obtained is
\begin{eqnarray}
M = \frac{N^2}{4\pi^2(7-p) d_{p}\lambda^2}
\Bigl(
(8-p)U_0^{7-p}+(7-p)\alpha_p d_p \lambda
\Bigr).
\label{ADM_mass}
\end{eqnarray}
If we set $U_0=0$, we have the zero temperature limit
$M_0 = \frac{N^2}{4\pi^2\lambda}$, which is nothing but the 
tension of the $N$ D$p$-branes and  
corresponds to the disk diagram depicted in figure~\ref{tube1}. 
The rest $M-M_0$ comes from the vacuum amplitudes 
of open strings. 
The ``mixed'' diagrams do not contribute to the ADM mass, 
because they are of order $\mathcal{O}(\frac{1}{U^{2(7-p)}})$ 
as in the middle diagram in figure \ref{tube2}.

We can directly show the equivalence of the ADM mass and
the energy of the open string system.
In general, the ADM mass is defined as the source of 
the graviton in the weak field expansion.
As we have seen in figure \ref{tube1} and  \ref{tube2},
in our case, it is nothing but the expectation value of the graviton
emission vertex. There might be a doubt about the use of the weak field 
expansion near the D$p$-branes, since the space-time is
curved there. 
However, this treatment is correct for the calculation of the ADM mass, 
because what we need are processes of
single graviton emission as the right 
diagram of figure~\ref{tube2}. 
Then the coupling of the graviton to the
open strings in the leading order of the weak field expansion
is obtained
by setting the target space metric to
\begin{eqnarray}
 G_{\mu\nu}=\delta_{\mu\nu}+h_{\mu\nu},
\end{eqnarray}
in the worldsheet action of the string.
Therefore the source $T_{\mu\nu}$ of the graviton in (\ref{TmnStringFrame})
is equal to the energy momentum tensor of the open strings on the flat
D$p$-branes.

\subsection{Low Temperature Limit - Gauge/Gravity Correspondence}
As we have discussed, it is not clear whether the event
horizon is really formed or not. However, if $p\le 3$ and the 
temperature is low enough, we have a rather strong evidence for 
its formation.
In the low temperature limit, the 
open strings on the D$p$-branes can be described by SYM.
Since SYM is finite for $p\le 3$,
the energy density is given by dimensional analysis as
\begin{eqnarray}
M=M_0+\beta^{-p-1}f(\lambda \beta^{3-p}).
\label{energy1 SYM}
\end{eqnarray}

On the gravity side, if we assume the existence of 
the event horizon, we can determine 
$U_0$ as a function of $\beta$ by using (\ref{inverse_hawking_temp}) 
and then the energy density using (\ref{ADM_mass}).
First, we consider the near extremal region
\begin{eqnarray}
U_0^{7-p} \ll d_p\lambda,
\label{near extremal limit 1}
\end{eqnarray}
which is equivalent to
\begin{eqnarray}
\beta\gg \lambda^{\frac{1}{7-p}},
\label{near extremal limit 2}
\end{eqnarray}
as long as (\ref{inverse_hawking_temp}) is satisfied.
Then (\ref{inverse_hawking_temp}) is solved as
$U_0=const.\times\left(\frac{\lambda}{\beta^2}\right)^{\frac{1}{5-p}}$
and (\ref{ADM_mass}) gives
\begin{eqnarray}
M=M_0+const.\times \beta^{-p-1}\left(\lambda \beta^{3-p}\right)^{-\frac{3-p}{5-p}},
\label{energy2 SYM}
\end{eqnarray}
which has the form of (\ref{energy1 SYM}).
Thus we have seen that the consequences of two assumptions
agree in a non-trivial manner.
One is that the open strings form the event horizon, 
and the other is that they are described by 
SYM in the low temperature region.
Therefore it is natural to conclude that both assumptions
are correct.

On the other hand, if the condition (\ref{near extremal limit 2}) 
is not satisfied, the energy density obtained from (\ref{inverse_hawking_temp}) 
and (\ref{ADM_mass}) no longer satisfies (\ref{energy1 SYM}).
Therefore either of the two assumptions breaks down.

\subsection{Open String/Gravity Correspondence}
In this section, we assume that the event horizon is
always there for any temperature.
Then we can trust (\ref{inverse_hawking_temp}) for
any $\beta$ and regard it as expressing 
the duality between the open string system
that is not necessarily reduced to SYM  
and the classical gravity.

To make the situation clearer, we rewrite 
 (\ref{inverse_hawking_temp}) in terms of
 scaling variables
\begin{eqnarray}
b = \frac{4\pi u_0}{7-p}
\left(
\frac{1}{2}+\frac{1}{2}\sqrt{1+\left(\frac{1}{u_0^{7-p}}\right)^{2}}
\right)^{\frac{1}{2}},
\label{b_definition}
\end{eqnarray} 
where 
\begin{eqnarray}
b &=& \frac{\beta}{(2d_p\lambda)^{\frac{1}{7-p}}}, 
\label{small_b}\\
u_0 &=& \frac{U_0}{(2d_p\lambda)^{\frac{1}{7-p}}}.  
\label{small_u0}
\end{eqnarray}

\begin{figure}[t]
\begin{center}
\includegraphics[scale=2.5]
                {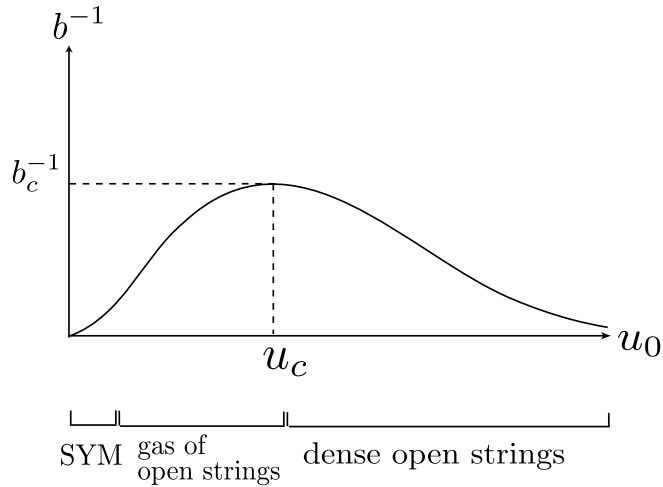}
\caption{Black $p$-brane temperature as a function of the horizon 
radius in the case $p<5$}
\label{nonextremal}
\end{center}
\end{figure}

As is shown in figure \ref{nonextremal}, two values of $u_0$ 
correspond to the same temperature, if $p<5$. 
If $u_0\ll 1$, the system is described by SYM and 
the temperature behaves as $b^{-1}\sim u_0^{\frac{5-p}{2}}$. 
If we keep increasing the energy, the horizon 
radius $u_0$ becomes larger, while the temperature $b^{-1}$ 
becomes maximum $b_c^{-1}$ at $u_0=u_c$, and then 
decreases to zero.

Let us see what happens to the system of open strings 
when the energy is increased from zero to infinity.
At the beginning when $u_0\ll 1$, the open 
strings are in a low energy state and described by SYM. 
When we increase the energy, stringy excitations become 
important and the system is no longer described by SYM, 
but still we can imagine it as a gas of open strings. 
However, this picture breaks down when the temperature
reaches the maximum value $b_c^{-1}$,
beyond which the heat capacity 
becomes negative. 
Although this transition is reminiscent of 
the Hagedorn transition, some differences are there. 
Firstly, this transition is caused by the interaction 
between the open strings. In fact, the maximum temperature 
$T_{max}$ becomes lower as $\lambda$ is increased as 
is seen from (\ref{small_u0}): 
\begin{eqnarray}
T_{max} = \frac{1}{b_c (2d_p\lambda)^{\frac{1}{7-p}}}.
\end{eqnarray}  
Secondly the negative heat capacity indicates that the density 
of states becomes even larger if we further increase 
the energy, which is contrary to what is 
expected from the Hagedorn transition \cite{AW88}. 
We finally comment that the metric (\ref{metric_finite}) becomes 
the Schwarzschild metric for large values of $u_0$,  
so that Schwarzschild black hole is realized as a system 
of dense and strongly coupled open strings.

\section{Wilson Loop and Minimal Surface}
\label{wilson_loop}
In this section, we discuss the correspondence of Wilson loops. 
The procedure is analogous to that in \cite{KS07} and 
we use a worldsheet symmetry to relate SYM side and 
the gravity side directly. 
For a while, we discuss the zero temperature case.

To find the symmetry, we examine
the black $p$-brane solution at zero temperature,
which is written as (\ref{metric_finite}) with 
$U_0$ set to zero.  
In the near horizon region 
\begin{eqnarray}
U^{7-p} \ll d_p\lambda,
\label{near_horizon_limit}
\end{eqnarray} 
it becomes
\begin{eqnarray}
ds^2=\Bigl(\frac{U^{7-p}}{d_{p}\lambda}\Bigr)^{\frac{1}{2}}
\Bigl[dx_{p+1}^2+
\displaystyle{\sum_{a=1}^{p}}dx_a^2\Bigr]+
\Bigl(\frac{d_{p}\lambda}{U^{7-p}}\Bigr)^{\frac{1}{2}}
\Bigl[dU^2+U^2d\Omega_{8-p}^2\Bigr].
\label{metric_nearhorizon}
\end{eqnarray}
This metric is 
invariant under the generalized scale transformation
\begin{eqnarray}
x_A\ &\to& c^{-1}x_A \quad (A=1,2,\cdots,p+1),\label{transf_x} \\
U,\ &\to& cU, \\
\lambda\ &\to& c^{3-p} \lambda. \label{transf_lambda},
\end{eqnarray}
where $c$ is an arbitrary real positive number.
Here we have checked the symmetry in the supergravity limit.
However, as we will see below,
it holds in the stringy level 
provided that the worldsheet is planar and located within 
the near horizon region and that $\lambda\gg 1$.

\begin{figure}[t]
\begin{center}
\includegraphics[scale=3.0]
                {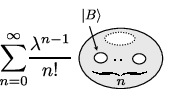}
\caption{A Wilson loop in terms of the worldsheets.}
\label{kappa}
\end{center}
\end{figure}

The Wilson loop is given by a power series of $\lambda$ as 
\begin{eqnarray}
W(\lambda) =\displaystyle{\sum_{n=0}^{\infty}}
\frac{\lambda^{n-1}}{n!}W_n.
\label{expansion_wilson}
\end{eqnarray}
Here $W_n$ is the contribution of the worldsheet that has $n$ boundaries 
corresponding to the coincident $N$ D$p$-branes and one boundary corresponding  
to the source D-brane as depicted in figure~\ref{kappa}.
At each solid circle we impose the boundary condition that
corresponds to the D$p$-brane, and give factor $N$.
At the dashed circle we take the boundary condition for
the source D-brane which creates the Wilson loop.

Before showing the invariance of the Wilson loop,
we mention how the gauge/gravity correspondence
is shown by this symmetry.
We start with a configuration in which the source
D-brane is widely extended and located close to the D$p$-branes,
as shown in the left picture in figure~\ref{wilsontrans}.
In this case the Wilson loop is described by the low
energy effective theory of the open strings, that is, SYM.
Then we apply the generalized scale transformation 
so that the worldsheet is stretched in the vertical direction
and shrunk in the horizontal direction,
as in the middle picture in figure~\ref{wilsontrans}. 
Then, as in the right picture, 
the amplitude can be evaluated as the area of the minimal surface 
in the black $p$-brane background. 
If we locate the source D-brane sufficiently close to the D$p$-branes
in the original configuration, 
we can keep it in the near horizon region
during the deformation.
In this way, the gauge/gravity correspondence follows simply,
once we can show the invariance.

\subsection{Scale Transformation of Boundaries}
Let us examine how the boundaries transform under the scale transformation
\footnote{In the following discussion, we consider type IIB superstring 
for concreteness but 
generalization to type IIA is straightforward. 
}.
In the Green-Schwarz formalism, 
the boundary state of the D$p$-branes is given by
\begin{eqnarray}
|B(p,X_i)\rangle
&=&\exp\Biggl[\displaystyle{\sum_{n=1}^{\infty}}
\Bigl(\frac{1}{n}M^{ij}\alpha^i_{-n}\tilde{\alpha}^j_{-n}
-iM^{ab}S^{a}_{-n}\tilde{S}^{b}_{-n}\Bigr)\Biggr]|B_0\rangle,
\label{boundary_state}\\
|B_0\rangle &=& 
C|p_A=0\rangle|x_i=X_i\rangle
\Bigl(M^{ij}| i\rangle | \tilde{j}\rangle-M^{\dot{a}\dot{b}}
|\dot{a}\rangle|\dot{b}\rangle\Bigr)\displaystyle
{\bigotimes_{n=1}^{\infty}}|0_n \rangle
\label{boundary_state_zero},
\end{eqnarray} 
where $C$ is a constant and 
$X_i$ is the location of the D$p$-branes \cite{GG96}. 
$M^{ij}$, $M^{ab}$ and $M^{\dot{a}\dot{b}}$ 
are defined by 
\begin{eqnarray} 
M^{ij} &=&
\left(
  \begin{array}{cc}
     -\mathbf{1}_{p+1}  & 0   \\
     0  &  \mathbf{1}_{7-p}  \\
  \end{array}
\right), \\
M^{ab}&=&(\gamma_1\cdots\gamma_{p+1})^{ab},
\quad M^{\dot{a}\dot{b}}=(\gamma_1\cdots\gamma_{p+1})^{\dot{a}\dot{b}}.
\end{eqnarray}

Under the infinitesimal scale transformation
$c=1+\epsilon$, 
worldsheet coordinates, their conjugate momenta and their super partners 
transform as
\begin{eqnarray}
\delta X^i(\sigma) = \epsilon M^{ij}X^{j},
\quad \delta P^i(\sigma) = -\epsilon M^{ij}P^{j}, \\
\delta S^a(\sigma) =i\epsilon M^{ab}\tilde{S}^b,
\quad
\delta \tilde{S}^a(\sigma) =-i\epsilon M^{ab}S^b,
\end{eqnarray}
or, in terms of the modes we have
\begin{eqnarray}
\delta \alpha_n^i =-\epsilon M^{ij}\tilde{\alpha}_{-n}^j, 
\quad \delta\tilde{\alpha}_n^i =-\epsilon M^{ij}\alpha_{-n}^j,\label{scaleTrans_X}\\ 
\delta S_n^a = i\epsilon M^{ab}\tilde{S}_{-n}^{b}, 
\quad 
\delta \tilde{S}_n^{a} =-i\epsilon M^{ab}S_{-n}^b.\label{scaleTrans_S}
\end{eqnarray}

\begin{figure}[t]
\begin{center}
\includegraphics[scale=2.5]
                {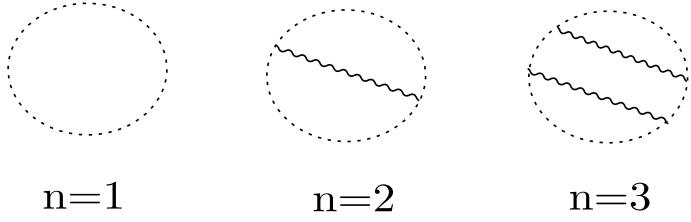}
\caption{Feynman diagrams of SYM corresponding to $W_n$. 
The dotted and wavy lines represent the contour of the Wilson loop and
a boson propagator respectively.}
\label{loops}
\end{center}
\end{figure}

The exponential factor of the RHS of (\ref{boundary_state}) is 
invariant under this transformation due to the supersymmetry,
and only the zero mode part (\ref{boundary_state_zero})
gives a nontrivial factor.
In order to determine this factor, we consider
the low energy limit, that is, SYM.
By a simple power counting we can show that
the Feynman diagrams of SYM corresponding to $W_n$
are proportional to $L^{(3-p)(n-1)}$, where $L$
is the size of the Wilson loop (see figure~\ref{loops})\footnote{
The first term $W_0$ has no counterpart of SYM. However  
it is negligible compared to the other terms, if 
the source D-brane is placed sufficiently close to the D$p$-branes. 
Then $W_0$ obeys the area law, while the other $W_n$'s do the perimeter
law. }. 
For example, in the SYM limit, $W_1$ becomes $1$, and $W_2$
represents one-boson exchange and is of 
order $L^{(3-p)}$.
Because $L$ transforms as $x_A$ (\ref{transf_x}),
we find that the contribution to the scale 
transformation coming from the boundaries
is given by $c^{-(3-p)(n-1)}$.

Therefore if we apply the scale transformation 
(\ref{scaleTrans_X}) and (\ref{scaleTrans_S}) to
the Wilson loop (\ref{expansion_wilson}),
the factor coming from the boundaries is exactly 
cancelled by that from $\lambda^{n-1}$, provided
that $\lambda$ is transformed as (\ref{transf_lambda}).
Then the variation of the Wilson loop simply
comes from that of the worldsheet action.

\subsection{Scale Transformation of  Wilson Loops}
\begin{figure}[t]
\begin{center}
\includegraphics[scale=2.5]
                {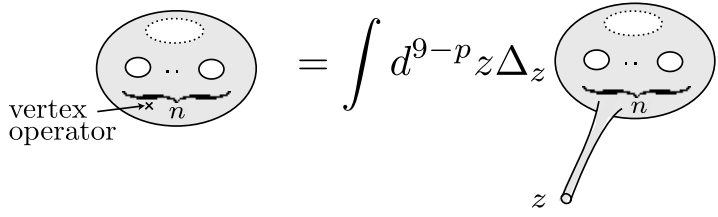}
\caption{A pictorial description of LSZ-like method.}
\label{lsz}
\end{center}
\end{figure}

The variation of the worldsheet action
is expressed as an insertion of the vertex operator that
corresponds to the level zero part $|B_0\rangle$ 
of the boundary state of the  D$p$-branes \cite{KS07}.
In order to evaluate such insertion, 
we introduce an extra D$p$-brane that is  
placed at $X_i=z_i$ $(i=p+2,\cdots,10)$. 
By $W_{n+1}(z)$ we denote the worldsheet of $W_n$
with one extra boundary added that correspond to the D$p$-brane 
placed at $X_i=z_i$ (see the right picture of figure \ref{lsz}).
Then the operator insertion is expressed by 
an LSZ-like formula \cite{KS07}
\begin{eqnarray}
\delta_{S} W_n = \epsilon\int d^{9-p}z \Delta_z W_{n+1}(z).
\end{eqnarray} 
Here $\delta_S$ stands for the variation coming 
from that of the worldsheet action.
By summing up for $n$, we have
\begin{eqnarray}
\delta_S W(\lambda) &=& \epsilon\int d^{9-p} z \Delta_z W(\lambda,z), 
\label{devs_int}\\
W(\lambda,z) &=& \displaystyle{\sum_{n=0} ^{\infty}} 
\frac{\lambda^{n-1}}{n!} W_{n+1}(z).  
\end{eqnarray}
Suppose that the source D-brane is located at 
a distance $h$ from the $N$ D$p$-branes. 
Then the worldsheet is stretched approximately 
in a region 
$0\le |X_i| \le U$, where $U=\text{max}(l_s,h)$. 
For $0\le |z_i| \le U$, we assume that $W(\lambda, z)$ varies 
slowly with $|z_i|$, that is, 
\begin{eqnarray}
W(\lambda,z)\sim W(\lambda,0).
\end{eqnarray}
When $z_i$ is larger than $U$ but inside the near horizon region, 
the tube stretching to the boundary at $X_i=z_i$ becomes thin and 
it is described by a propagation of massless states of closed string.
Therefore 
\begin{eqnarray}
W(\lambda,z)\sim \frac{U^{7-p}}{z^{7-p}}W(\lambda, z=U)
\sim \frac{U^{7-p}}{z^{7-p}}W(\lambda, z=0),
\end{eqnarray} 
for $|z_i|\ge U$.

From the last estimate, (\ref{devs_int}) is evaluated as
\begin{eqnarray}
\delta_{S} W (\lambda) \sim \epsilon U^{7-p} W(\lambda, z=0 )
=\frac{\epsilon U^{7-p}}{\lambda}\partial_{\lambda} (\lambda W(\lambda)).
\end{eqnarray}
Here we have used the identity
\begin{eqnarray}
W(\lambda,z=0) = \frac{\partial_{\lambda}(\lambda W(\lambda))}{\lambda},
\end{eqnarray}
which follows from $W_{n+1}(z=0)=W_{n+1}$.
Therefore we obtain 
\begin{eqnarray}
\delta_S \log (\lambda W(\lambda)) \sim
\epsilon U^{7-p}\partial_{\lambda}\log(\lambda W(\lambda))
< \frac{\epsilon U^{7-p}}{\lambda} \log (\lambda W(\lambda)),
\end{eqnarray}
which indicates $\delta_SW(\lambda)$ is negligible compared to the 
Wilson loop $W(\lambda)$ itself, if the source D-brane is
located in the near horizon region $U^{7-p}\ll\lambda$.
Here we have assumed that $\log (\lambda W(\lambda))$ is bounded by a polynomial
for large $\lambda$ so that 
$\partial_{\lambda} \log (\lambda W(\lambda))$ is bounded by 
$\log(\lambda W(\lambda))/\lambda$.
  
\subsection{Generalization to Finite Temperature}
Generalization to the case of finite temperature is straightforward.  
In the supergravity limit, the black $p$-brane solution at finite temperature
is written as (\ref{metric_finite}) and  
in the near horizon region (\ref{near_horizon_limit})
it becomes 
\begin{eqnarray}
ds^2=\Bigl(\frac{U^{7-p}}{d_{p}\lambda}\Bigr)^{\frac{1}{2}}
\Bigl[\Bigl(1-\Bigl(\frac{U_0}{U}\Bigr)^{7-p}\Bigr)dx_{p+1}^2+
\displaystyle{\sum_{a=1}^{p}}dx_a^2\Bigr]+
\Bigl(\frac{d_{p}\lambda}{U^{7-p}}\Bigr)^{\frac{1}{2}}
\Bigl[\frac{dU^2}{1-(\frac{U_0}{U})^{7-p}}+U^2d\Omega_{8-p}^2\Bigr].
\nonumber\\
\label{metric_nearhorizon_finite}
\end{eqnarray}
Here we have also assumed the near extremal 
limit (\ref{near extremal limit 1})
in order that the event horizon is inside the near horizon region.
This metric is 
invariant under the generalized scale transformation
\begin{eqnarray}
x_A\ &\to& c^{-1}x_A \quad (A=1,2,\cdots,p,p+1), \\
U,\quad U_0 &\to& cU, \quad cU_0 \\
\lambda\ &\to& c^{3-p} \lambda,
\end{eqnarray}
where $c$ is an arbitrary real positive number.

We can show that this symmetry is extended to the
stringy level as in the zero-temperature case.
The boundary state of the D$p$-branes 
at finite temperature is given by \cite{Sen98},
\begin{eqnarray}
|B(p,X^i)\rangle
&=&\exp\Biggl[\displaystyle{\sum_{n=1}^{\infty}}
\Bigl(\frac{1}{n}M^{ij}\alpha^i_{-n}\tilde{\alpha}^j_{-n}
-iM^{ab}S^{a}_{-n}\tilde{S}^{b}_{-n}\Bigr)\Biggr]|B_0\rangle,\\
|B_0\rangle &=& 
C\beta\displaystyle{\sum_{w\in\mathbf{Z}}}e^{i\theta w}|w\rangle
|p_a=0,n_{KK}=0\rangle|x^i=X^i\rangle
\Bigl(M^{ij}|i\rangle | \tilde{j}\rangle-M^{\dot{a}\dot{b}}
|\dot{a}\rangle|\dot{b}\rangle\Bigr)\displaystyle
{\bigotimes_{n=1}^{\infty}}|0_n \rangle,
\nonumber\\
\end{eqnarray}
where $w$ is the winding number, $n_{KK}$ is the Kaluza-Klein momentum and 
$\theta$ is an expectation value of the Wilson line along the 
Euclidean time direction.  
Under the scale transformation, 
this state transforms in the same manner as 
that of zero temperature, and we can apply the
same analysis to show the invariance of
the Wilson loop as before.

Then by using the transformation,
we can relate the SYM side to the supergravity side.
The only thing we need to check is the validity of  
the classical supergravity, which we discuss below. 

\subsection{Validity of the Classical Gravity}
Using the generalized scale transformation,
we can show that the Wilson loop in SYM  
(the left in figure~\ref{wilsontrans}) is equivalent to 
the long-stretched worldsheet (the middle in figure~\ref{wilsontrans}). 
In this subsection we examine when the latter can be evaluated by 
classical gravity. 

For simplicity, we consider the Polyakov line here.
We assume that the source D-brane is placed sufficiently
distant from the event
horizon $h\gg U_0$, where $h$ is the distance between the source
D-brane and the D$p$-branes.
First, the conditions (\ref{limit_gravity}) must be satisfied
after the transformation in order for classical gravity to be applicable. 
Hence the scaling factor $c$ should satisfy
\begin{eqnarray}
c^{-1}\beta\gg 1,
\qquad
c^{3-p}\lambda\gg 1,  
\end{eqnarray} 
or equivalently,
\begin{eqnarray}
\lambda^{-1/(3-p)}
\ll
c
\ll 
\beta. 
\label{c and beta and lambda}
\end{eqnarray}
In order for $c$ to exist we need 
\begin{eqnarray}
\beta^{3-p}\lambda\gg1.\label{strong_eff_coupling}
\end{eqnarray} 
Note that $\beta^{3-p}\lambda$ is invariant under 
the generalized scale transformation, and can be regarded as the
effective coupling constant of SYM as in (\ref{energy1 SYM}). 

Next, we require that after the transformation
the source D-brane is located sufficiently far from
the D$p$-branes so that
the stringy excitations are suppressed around the minimal
surface.  
At the same time it must be in the near horizon region 
in order for the generalized scale transformation to be applicable. 
Therefore, we need  
\begin{eqnarray}
1\ll ch, 
\qquad
ch\ll (c^{3-p}\lambda)^{1/(7-p)},   
\label{height_vs_string}
\end{eqnarray}
or equivalently, 
\begin{eqnarray}
h^{-1}
\ll
c
\ll
\lambda^{1/4}h^{-(7-p)/4}. 
\label{constraint_height}
\end{eqnarray}  

In summary, the classical gravity is valid when 
(\ref{c and beta and lambda}) and 
(\ref{constraint_height}) are satisfied. It is easy 
to check that such $h$ and $c$ exist if (\ref{strong_eff_coupling})
is satisfied, assuming $p\le 3$.
Then the right picture in figure \ref{wilsontrans} can be evaluated 
as the area of the minimal surface in the black $p$-brane background.   

\section{Conclusions and Discussions}
\label{conclusions_and_discussions}

In this paper, we investigated gauge/gravity 
dualities from the worldsheet viewpoint. We directly 
related SYM to supergravity by analyzing worldsheets. 

For the energy density or entropy, we have considered 
open strings on the D-branes and a single graviton 
emission from them. 
We identified the graviton exchange amplitude with the 
leading deviation of the metric from the flat one.
Since the graviton is directly coupled to the energy momentum tensor of
the open strings, we can identify  
the ADM mass per unit volume with the energy density of the open strings.  
In this discussion, no symmetry is necessary to 
show the correspondence. We can also consider the non-extremal case since 
the near horizon limit is not needed for our discussion.
It is not clear whether an event horizon is formed or not. 
If it is the case, a non-extremal black brane is realized as 
a system of dense and strongly coupled open strings. 
It would be interesting to investigate this system in detail. 

For Wilson loops, on the other hand, 
we take the near horizon limit in order to  
use the generalized scale invariance, by which we can
relate the Wilson loops in SYM to the minimal surfaces in the 
classical curved background.  
This analysis can be easily generalized to the GKPW relation. 
It is also interesting to apply our argument to other systems, 
although it is limited to systems which can be embedded in string theory.

In this paper, we have mainly discussed the correspondence 
between SYM and classical gravity. We can also consider 
$\alpha'$-corrections by taking $\lambda$ 
large but finite. In SYM, 
they correspond to $1/\lambda^{\frac{2}{7-p}}$ expansion.
For the energy density, we can repeat the same argument as section 2.
We can still identify the ADM mass with the energy density
of the D$p$-branes. 
The only difference is that we should include the
higher derivative corrections to the supergravity equation of motion.
Then the black $p$-brane solution is modified, 
and the relations among $M$, $U_0$, $\beta$ receive corrections,
but the rest is the same.

For the Wilson loop, we can consider the $\alpha'$-corrections as follows.
This time, we need to consider not only the minimal surface but also 
the fluctuations of the non-linear sigma model  
\begin{eqnarray}
S=-\frac{\lambda^{\frac{2}{7-p}}}{2\pi}
\int_{\Sigma} d^2\sigma G_{\mu\nu}(X)\partial_a X^{\mu}\partial^a X^{\nu}+\cdots.
\label{nlsigma}
\end{eqnarray}
Here $G_{\mu\nu}$ is the black $p$-brane metric with $\alpha'$ corrections, 
and $\Sigma$ is a disk whose boundary is on the source D-brane.
When $\lambda$ is finite, the near horizon region is not infinitely large. 
Therefore the best we can do by the generalized scale transformation is to 
lift the source D-brane to a distance of order $\lambda^{\frac{1}{7-p}}$ 
from the D$p$-branes. Then the exchange of massive modes
between the D$p$-branes and $\Sigma$ is not completely negligible, 
but is of order $\mathcal{O}(\exp(-\lambda^{\frac{1}{7-p}}))$. 
However, as long as $\lambda$ is large compared to 1,
this is negligible compared to the perturbative effects in (\ref{nlsigma}),
which are expressed as a power series in $1/\lambda^{\frac{2}{7-p}}$. 
Thus we can justify that the $\alpha'$ corrections of the classical 
string corresponds to the $1/\lambda^{\frac{2}{7-p}}$ expansion of SYM.

\begin{figure}[t]
\begin{center}
\includegraphics[scale=2.0]
                {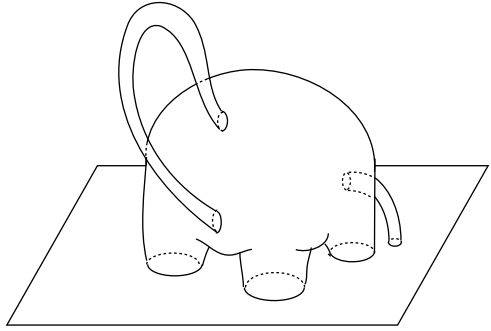}
\caption{A worldsheet with a handle, which can be stretched 
outside the near horizon region. 
}
\label{fig:handle}
\end{center}
\end{figure}

Though we can include $\alpha'$ corrections in this manner,  
it seems to be very difficult to include quantum effects of closed strings, 
or in other words, to take $N$ to be finite \cite{Park99}. 
If we allow the worldsheet to have a handle, it can be stretched 
outside the near horizon region, as shown in figure~\ref{fig:handle}, 
and the generalized scale transformation is not a symmetry any more \cite{KS07}.

\begin{center}
\noindent {\bf Acknowledgments}
\end{center}
This work was supported by the Grant-in-Aid for 
the Global COE Program ``The Next Generation of Physics, Spun from 
Universality and Emergence'' from the Ministry of Education, Culture, 
Sports, Science and Technology (MEXT) of Japan.
T.~A.~ is supported by the Japan Society for the Promotion of Science.
M.~H. would like to thank Y.~Hyakutake, A.~Miwa and J.~Nishimura for 
discussions on related works.

\end{document}